\documentstyle[aps,pra,epsfig,preprint]{revtex} 
\tightenlines

\def\be{\begin{equation}}
\def\ee{\end{equation}}
\def\bea{\begin{eqnarray}}
\def\eea{\end{eqnarray}}
\def\bma{\begin{mathletters}}
\def\ema{\end{mathletters}}
\def\C{\hbox{$\mit I$\kern-.6em$\mit C$}}

\begin{document}
\draft

\title{The distillable entanglement of multiple copies of Bell states}

\author{Yi-Xin Chen and Dong Yang}

\address{Zhejiang Institute of Modern Physics and
Department of Physics, Zhejiang University, Hangzhou 310027, P.R. China} 

\date{\today}

\maketitle

\begin{abstract}
It is impossible to discriminate four Bell states  through local operations and classical communication (LOCC), if only one copy is provided. To complete this task, two copies will suffice and be necessary. When $n$ copies are provided, we show that the distillable entanglement is exactly $n-2$.
\end{abstract}


\narrowtext

The no-cloning theorem \cite{Wootters} asserts that it is impossible to discriminate  non-orthogonal states with certainty. In general, orthogonal states may be distinguished perfectly only by means of global measurements since quantum information of orthogonality may be encoded in entanglement which may not be extracted by LOCC operations. If only LOCC operations are allowed, even product orthogonal states could not be discriminated exactly \cite{Bennett}.  However, Walgate et al. \cite{Walgate1} demonstrated that any two orthogonal multipartite states could be discriminated with certainty by only LOCC operations. To discriminate multiple orthogonal states, more copies are required. They also showed that $n$ possible orthogonal states can be distinguished perfectly with $n-1$ copies. It is an upper bound upon the number of copies required for local distinction of states. Further, they pointed out that there are sets of orthogonal states that can be distinguished using less than $n-1$ copies. In the case of four Bell states, two copies will suffice. Recently, using the existing inequality among the measures of entanglement, Ghosh et al. \cite{Ghosh} proved that any three Bell states cannot be discriminated by LOCC operations. From a different point of view, Walgate and Hardy \cite{Walgate2} arrived at the same conclusion and discussed the sufficient and necessary conditions to discriminate $2\times n$ states. The question of local distinction of non-orthogonal states has also been investigated in recent papers \cite{Virmani,CY}. In \cite{Ghosh}, Ghosh et al. calculated the distillable entanglement \cite{Bennett1} of the mixed state comprising of two of the Bell basis with equal a priori probability. In this note, we prove that if $n$ copies out of four Bell states are provided, the distillable entanglement is $n-2$. \\

Denote the four Bell states as
\bea
|\Phi_{1} \rangle=\frac{1}{\sqrt{2}}(|00\rangle+|11\rangle), \nonumber \\
|\Phi_{2} \rangle=\frac{1}{\sqrt{2}}(|00\rangle-|11\rangle), \nonumber \\
|\Phi_{3} \rangle=\frac{1}{\sqrt{2}}(|01\rangle +|10\rangle), \nonumber \\
|\Phi_{4} \rangle=\frac{1}{\sqrt{2}}(|01\rangle-|10\rangle).
\eea

In \cite{Ghosh}, the distillable entanglement of the mixed state of the form 
\be
\rho=\frac{1}{2}(|\Phi_{i}\rangle^{\otimes 2}\langle\Phi_{i}|+\Phi_{j}\rangle^{\otimes 2}\langle\Phi_{j}|),
\ee
is shown to be one ebit, where $i\neq j$. 

In the following, we will consider the distillable entanglement of the mixed state comprising of four Bell states 
\be
\rho^{(n)}=\frac{1}{4}\sum_{i=1}^{4}|\Phi_{i}\rangle^{\otimes n}\langle\Phi_{i}|.
\ee 

For $n=1$, it is explicit that $\rho^{(1)}$ is separable, so $E_{d}(\rho^{(1)})=0$. 

For $n=2$, $\rho^{(2)}$ is also separable \cite{Smolin}, so $E_{d}(\rho^{(2)})=0$. Though the proof of $\rho^{(2)}$ is simple, it is of importance to calculate the relative entanglement \cite{VP} of the mixed state of the form Eq(3) for $n=2m$. 

Recall that the relative entanglement of mixed state $\rho$ is defined as 
\be
E_{r}(\rho)=min_{\sigma\in\cal{D}}S(\rho\|\sigma),
\ee 
where $\cal{D}$ is the set of separable states, $S(\rho\|\sigma)=Tr\rho(\log\rho -\log\sigma)$ is the relative entropy. We will show that $\rho^{(2)\otimes m}$ minimizes of  $S(\rho^{(2m)}\|\sigma)$ over $\sigma\in\cal{D}$. Since $\rho^{(2)}$ is separable, $\rho^{(2)\otimes m}$ is also separable. By straight computation, 
\bea
S(\rho^{(2m)}\|\rho^{(2)\otimes m})&=&Tr\rho^{(2m)}(\log\rho^{(2m)}-\log\rho^{(2)\otimes m}), \nonumber\\
&=&-2-\frac{1}{4}\sum_{i=1}^{4}Tr[|\Phi_{i}\rangle^{\otimes 2m}\langle\Phi_{i}|\log(2^{-2m}\sum_{k_{1},k_{2},\cdots k_{m}=1}^{4}\otimes_{j=1}^{m}|\Phi_{k_{j}}\rangle^{\otimes 2}\langle\Phi_{k_{j}}|)], \nonumber\\
&=&-2+2m\times \frac{1}{4}\sum_{i=1}^{4}\langle\Phi_{i}^{\otimes 2m}|\sum_{k_{1},k_{2},\cdots k_{m}=1}^{4}\otimes_{j=1}^{m}|\Phi_{k_{j}}\rangle^{\otimes 2}\langle\Phi_{k_{j}}|)|\Phi_{i}\rangle^{\otimes 2m}, \nonumber\\
&=&2m-2.
\eea
So it is easy to know
\be
E_{r}(\rho^{(2m)})\le S(\rho^{(2m)}\|\rho^{(2)\otimes m})=2m-2.
\ee   
On the other hand, we know that the relative entanglement is an upper bound on the distillable entanglement, that is 
\be
E_{d}(\rho^{(2m)})\le E_{r}(\rho^{(2m)}).
\ee
Further, the distillable entanglement is the maximal number of arbitrarily pure singlets that can be prepared locally from mixed state by entanglement purification protocols, here by LOCC operations. So the entanglement distilled by any definite protocol is not larger than the distillable entanglement. In \cite{Walgate1}, it was showed that two copies suffice to distinguish the four Bell states. We employ the distinction process for distillation of entanglement and at least $2m-2$ ebits could be obtained since only two copies are discarded. So we have the inequality
\be
2m-2\le E_{d}(\rho^{(2m)}).
\ee
Now it is clear that for $n=2m$
\be
E_{d}(\rho^{(2m)})= E_{r}(\rho^{(2m)})=2m-2.
\ee

For $n=2m+1$, we have not found the optimal separable state $\sigma$ which minimizes $S(\rho^{(2m+1)}\|\sigma)$ over $\sigma\in\cal{D}$, so $E_{r}(\rho^{(2m+1)})$ is unknown. However, we can argue that $E_{d}=n-2$ is also true by the same reasoning, 
\bea
S(\rho^{(2m+1)\otimes 2}\|\rho^{(2)\otimes 2m+1})=4m-2, \nonumber\\ 
E_{r}(\rho^{(2m+1)}) \le \frac{1}{2}S(\rho^{(2m+1)\otimes 2}\|\rho^{(2)\otimes 2m+1}), \nonumber\\
2m-1 \le E_{d}(\rho^{(2m+1)}) \le E_{r}(\rho^{(2m+1)}).
\eea
So $E_{d}(\rho^{(n)})=n-2$ also holds when $n$ is odd. 

Clearly, it is easy to see that $\rho^{(2)\otimes n}$ realizes the minimization of $S(\rho^{(n)\otimes 2}\|\sigma)$ over $\sigma\in\cal{D}$. As a byproduct,  $E_{r}(\rho^{(n) \otimes 2})=2n-4$. When $n$ is even, $E_{r}(\rho^{(2m)})=2E_{r}(\rho^{(m)})$ and the additivity of the relative entanglement holds. Since  $E_{r}(\rho^{(2m+1)})$ is unknown, we do not know whether $E_{r}(\rho^{(2m+1) \otimes 2})$ is strictly larger than $2E_{r}(\rho^{(2m+1)})$. If that is the case, it is a counterexample \cite{VW} to the additivity conjecture for the relative entanglement. In addition, we can conclude from the above discussion that two copies are sufficient and necessary to discriminate the four Bell states by LOCC operations. If it is not true, more than $n-2$ ebits could be distilled out of the mixed state $\rho^{(n)}$ which contradicts our conclusion. 

Notice that all permutations of the four Bell states could be realized by only local unitary operations. It is sufficient to show that any two of them could be interchanged locally while the other two remain unchanged. For example, under the local unitary transformation $|0\rangle \rightarrow |0\rangle, |1\rangle \rightarrow e^{i\frac{\pi}{2}}|1\rangle$, $|\Phi_{1} \rangle\leftrightarrow|\Phi_{2} \rangle$ while $|\Phi_{3} \rangle $ and $|\Phi_{4} \rangle $ are unchanged ignoring the global phase. Similarly, the interchange between other states could be obtained locally. It is the particular property of Bell states. For the generalized Bell states in higher dimensionality, not all permutations of the bases could be transformed by only local unitary operations. Now we can further generalize our outcome to the mixed states of the form
\be
\sigma^{(n)}=\frac{1}{4}\sum_{i_{j}=1}^{4}\otimes _{j=1}^{n}|\Phi_{i_{j}}\rangle\langle\Phi_{i_{j}}|,
\ee 
where there is only four terms in the sum and the corresponding states of the four terms form a permutation of the Bell bases. Through local unitary operations, $\sigma^{(n)}$ could be transformed to $\rho^{(n)}$. So the distillable entanglement of $\sigma^{(n)}$ is also $n-2$. \\

As well known, entanglement is responsible for many quantum tasks and pure entangled states are required in most cases. Unfortunately, entanglement is fragile and easy to be blurred by noise, so distillation of entanglement is of importance. Though many distillation protocols and upper bounds are known, distillable entanglement are known in few nontrivial cases. In this note, we have shown that when $n$ copies out of four possible Bell states are provided with equal a priori probability, the distillable entanglement is exactly $n-2$. 
\\

D. Yang thanks S. J. Gu and H. W. Wang for helpful discussion. We also thank Aditi Sen(DE) and Ujjwal Sen for commenting on the paper. The work is supported by the NNSF of China, the Special NSF of Zhejiang Province (Grant No.RC98022) and Guang-Biao Cao Foundation in Zhejiang University.


\end{document}